\documentclass[prl,twocolumn,superscriptaddress,showpacs,preprintnumbers,amsmath,amssymb]{revtex4}
\usepackage{mathrsfs}
\usepackage{bbm}
\usepackage{amsfonts}
\usepackage{tipa}

\usepackage{epsfig,graphicx}
\usepackage{amstext}
\usepackage{amsmath}
\usepackage{graphicx}
\usepackage{times}

\begin{document}
%%%%%%%%%%%%%%%%%%%%%%%%%%%%%%%%%%%%%%%%%%%%%%%%%%%%%%%%%%%%%%%%%%%%%%

%TCIDATA{OutputFilter=Latex.dll}
%TCIDATA{Version=5.00.0.2552}
%TCIDATA{<META NAME="SaveForMode" CONTENT="1">}
%TCIDATA{LastRevised=Wednesday, June 22, 2005 16:21:09}
%TCIDATA{<META NAME="GraphicsSave" CONTENT="32">}

\title{Evolution equation for quantum coherence}

\author{Ming-Liang Hu}
\email{mingliang0301@163.com}
\affiliation{School of Science, Xi'an University of Posts and Telecommunications,
             Xi'an 710121, China}
\author{Heng Fan}
\email{hfan@iphy.ac.cn}
\affiliation{Beijing National Laboratory for Condensed Matter Physics,
             Institute of Physics, Chinese Academy of Sciences, Beijing 100190, China}
\affiliation{Collaborative Innovation Center of Quantum Matter, Beijing 100190, China}

\begin{abstract}
The estimation of the decoherence process of an open quantum system
is of both theoretical significance and experimental appealing.
Practically, the decoherence can be easily estimated if the coherence
evolution satisfies some simple relations. Based on the coherence
quantification method, we prove a simple factorization relation for
the $l_1$ norm measure of coherence, and analyze under which condition
this relation holds. We also obtain a more general relation which
applies to arbitrary $N$-qubit state, and determine a condition for
the transformation matrix of the quantum channel which can support
permanently freezing of the $l_1$ norm of coherence. These results
simplify determination of a general decoherence dynamics to that the
investigation of evolution about the representative probe state.
\end{abstract}


\maketitle

\emph{Introduction.}--- Quantum coherence, an embodiment of the
superposition principle of states, lies at the heart of quantum
mechanics, and is also a major concern of quantum optics
\cite{Ficek}. Physically, coherence constitutes the essence of
quantum correlations (e.g., entanglement \cite{rmp-en} and quantum
discord \cite{rmp-qd}) in bipartite and multipartite systems which
are indispensable resources for quantum communication and
computation tasks. It also finds support in the promising subject
of thermodynamics \cite{ther1,ther2,ther3,ther4,ther5} and quantum
biology \cite{biolo}.

Due to the lack of rigorous coherence measures, researches in this
subject were usually limited to the analysis of the decay of the
off-diagonal elements of a density matrix, and only qualitative
statements are established. Sometimes, behaviors of
coherence were also analyzed indirectly via that of various quantum
correlations \cite{rmp-qd}. However, coherence and quantum correlations
are in fact different. Very recently, the characterization and
quantification of quantum coherence from a mathematically rigorous
and physically meaningful perspective has been achieved \cite{coher},
and this sets the stage for quantitative analysis of coherence. In the
past one year, researches in this field mainly center around the
establishment of various coherence monotones \cite{coher,meas1,meas2,
meas3,meas4,meas5} and their calculation \cite{bound}. Some other
progresses about coherence measures include the revelation of their
operational interpretation via entanglement \cite{meas2,inter1} and
discord-like correlations \cite{inter3,inter4,inter5}, their frozen in
noisy environments \cite{frozen}, their local and nonlocal creativity
\cite{create1, create2,create3}, their tradeoffs with other quantum
feature measures \cite{trade}, and the role they played in the
fundamental issue of quantum mechanics \cite{Pati,Guoyu, Winter}.

From a practical point of view, it is vital to make clear the
decoherence mechanism of a system when it is subject to the noisy
environments. The reason is twofold. First, the subject of
decoherence is a fundamental problem of modern physics, and
revealing its behavior can help to understand the subtle issue of
quantum mechanics from classical world \cite{Ficek}. Second,
coherence is a resource for quantum information and computation, but
the unavoidable interaction of quantum devices with the environment
often decoheres the input state and induces coherence loss, hence
damage the superiority of quantum information and computation
\cite{Nielsen}. Making clear dependence of the decoherence process
on structure of an environment can facilitate the design of
efficient coherence preservation protocols.

Looking for general law determining the decoherence process of a
system is of both theoretical significance and experimental
appealing. Remarkably, the evolution equations of certain
entanglement monotones (or their bounds) \cite{natphys,science,prl1,
prl2,Fac1,Fac2} and discord-like correlations \cite{hufan} were
found to obey the factorization relation for specific initial
states, and this simplifies the assessment of their robustness
against decoherence. Then, it is quite natural to ask whether there
exist similar relations for the coherence monotones. In this Letter,
we aimed at solving this problem. We first classify the general
$d$-dimensional states into different families, and then establish a
factorization relation which holds for them. By employing this
factorization relation, we further identified condition on the
quantum channels for freezing coherence. We also generalized this
factorization relation such that it applies to arbitrary $N$-qubit
state. The results are hoped to add another facet to the already
rich theory of decoherence, and shed light on revealing interplay
between structures of the quantum channel and geometry of the state
space, as well as how they determine quantum correlation behaviors
of an open system.

\emph{Coherence measures.}--- By establishing rigorously the sets of
incoherent states $\mathcal {I}$ which are diagonal in the chosen
reference basis $\{|i\rangle\}_{i=1,\dots,d}$, and incoherent
operations $\mathcal {E}_{\rm ICPTP}$ specified by the Kraus
operators $\{E_l\}$ which map all $\delta\in\mathcal {I}$ into
$\mathcal {I}$, Baumgratz and coworkers \cite{coher} proposed the
conditions for an information-theoretic coherence measure $C$: (1)
$C(\rho)\geq 0$ for all states $\rho$, and $C(\delta)=0$ iff
$\delta\in\mathcal {I}$. (2) Monotonicity under the actions of
$\mathcal {E}_{\rm ICPTP}$, $C(\rho)\geq C(\mathcal {E}_{\rm
ICPTP}(\rho))$. (3) Monotonicity under selective incoherent
operations on average, i.e., $C(\rho)\geq \sum_l p_l C(\rho_l)$,
where $\rho_l=E_l\rho E_l^\dag/p_l$, and $p_l={\rm Tr} (E_l\rho
E_l^\dag)$ is the probability of obtaining the outcome $l$. (4)
Convexity, $\sum_l p_l C(\rho_l) \geq C(\sum_l p_l \rho_l)$, with
$p_l \geq 0$ and $\sum_l p_l=1$.

There are several quantifiers which have been shown to be
well-defined coherence monotones. They are the $l_1$ norm, the
relative entropy \cite{coher}, the Uhlmann fidelity \cite{meas2},
and the intrinsic randomness \cite{meas4}. In this Letter, we
concentrate on the $l_1$ norm of coherence. For a given density
matrix $\rho$ and reference basis $\{|i\rangle\}_{i=1,\dots,d}$, it
is defined as \cite{coher}
%%%%%%%%%%%%%%%%%%%%%%%%%%%
\begin{eqnarray}\label{eq2-1}
 C(\rho) = \sum_{i\neq j} |\langle i| \rho |j\rangle|,
\end{eqnarray}
%%%%%%%%%%%%%%%%%%%%%%%%%%%
which equals the summation of the absolute values of the
off-diagonal elements of $\rho$.

\emph{General results.}--- Consider a general $d$-dimensional state
in the Hilbert space $\mathcal {H}$. The corresponding density
matrix can be written as
%%%%%%%%%%%%%%%%%%%%%%%%%%%
\begin{eqnarray}\label{eq3-1}
\rho = \frac{1}{d}\mathbb{I}_d+\frac{1}{2}\vec{x}\cdot\vec{X},
\end{eqnarray}
%%%%%%%%%%%%%%%%%%%%%%%%%%%
where $\mathbb{I}_d$ is the $d\times d$ identity matrix, $\vec{x}=
(x_1,x_2, \ldots, x_{d^2-1})$, with $x_i={\rm Tr}(\rho X_i)$. Here,
$\vec{X}=(X_1,X_2,\ldots, X_{d^2-1})$, with $X_i \varpropto T_i$,
and $\{T_i\}$ are generators of the Lie algebra ${\rm SU}(d)$
\cite{sud}. In the defining, or fundamental,  representation of
${\rm SU}(d)$, they are represented by the $d\times d$ traceless
Hermitian matrices, which satisfy $T_i T_j=\delta_{ij}\mathbb{I}_d
/2d+ \sum_{k=1}^{d^2-1} (i f_{ijk} + d_{ijk}) T_k /2$, where
$f_{ijk}$ are the structure constants that are completely
antisymmetric in all indices, while $d_{ijk}$ are completely
symmetric in all indices.

In the computational basis $\{|i\rangle\}_{i=1,\ldots,d}$, elements
of $\vec{X}$ can be written explicitly as
%%%%%%%%%%%%%%%%%%%%%%%%%%%
\begin{equation}\label{eq3-2}
 \begin{split}
  & u_{jk}= |j\rangle\langle k| + |k\rangle\langle j|, ~
    v_{jk}= -i(|j\rangle\langle k| - |k\rangle\langle j|), \\
  & w_{l}= \sqrt{\frac{2}{l(l+1)}}\sum_{j=1}^l (|j\rangle\langle j|
           - l |l+1\rangle\langle l+1|),
 \end{split}
\end{equation}
%%%%%%%%%%%%%%%%%%%%%%%%%%%
where $j,k\in \{1,2,\ldots, d\}$ with $j<k$, and $l\in \{1,2,\ldots,
d-1\}$. Clearly, $\{X_i\}$ satisfy ${\rm Tr} (X_i^\dag X_{i'}) =2
\delta_{ii'}$. Moreover, the notation $i$ appeared in $v_{jk}$ is
the imaginary unit.

For $\rho$ represented as Eq. \eqref{eq3-1}, the $l_1$ norm of
coherence can be derived as
%%%%%%%%%%%%%%%%%%%%%%%%%%%
\begin{eqnarray}\label{eq3-3}
 C_{l_1}(\rho)=\sum_{r=1}^{d_0} \sqrt{x_{2r-1}^2 + x_{2r}^2},
\end{eqnarray}
%%%%%%%%%%%%%%%%%%%%%%%%%%%
with $d_0=(d^2-d)/2$, and $\{X_i\}$ are arranged in the sequence:
$\vec{X}= \{u_{12},v_{12},\ldots, u_{d-1,d},v_{d-1,d}, w_1,\ldots,
w_{d-1}\}$. Clearly, $x_l$ related to $w_l$ which is diagonal in the
computational basis do not contribute to the $l_1$ norm of
coherence.

To investigate decay behaviors of the $l_1$ norm of coherence, we
suppose the considered system traverses a quantum channel $\mathcal
{E}$ specified by the set of Kraus operators $\{E_\mu\}$ satisfying
the completeness condition $\sum_\mu E_\mu^\dag E_\mu=\mathbb{I}_d$.
The evolution of this system can then be described by the completely
positive and trace preserving (CPTP) map \cite{Nielsen}
%%%%%%%%%%%%%%%%%%%%%%%%%%%
\begin{eqnarray}\label{eq3-4}
 \mathcal {E}(\rho)= \sum_\mu E_\mu \rho E_\mu^\dag
                   = \frac{1}{d}\mathbb{I}_d+\frac{1}{2}\vec{x}'\cdot\vec{X},
\end{eqnarray}
%%%%%%%%%%%%%%%%%%%%%%%%%%%
where elements of $\vec{x}'$ for $\mathcal{E}(\rho)$ are given by
$x'_i= {\rm Tr} [\mathcal{E}(\rho) X_i]$.

Now, we turn to the Heisenberg picture to describe $\mathcal{E}$ via
the map: $\mathcal{E}^\dag (X_i)= \sum_\mu E_\mu^\dag X_i E_\mu$. As
an arbitrary Hermitian operator $\mathcal{O}$ on $\mathbb{C}^{d
\times d}$ can be decomposed as $\mathcal{O}=\sum_{i=0}^{d^2-1} r_i
X_i$, ($r_i \in \mathbb{R}$), the map $\mathcal{E}^\dag (X_i)$ can
be further characterized by the transformation matrix $T$ defined
via
%%%%%%%%%%%%%%%%%%%%%%%%%%%
\begin{eqnarray}\label{eq3-5}
 \mathcal {E}^\dag (X_i)= \sum_{j=0}^{d^2-1} T_{ij}X_j,
\end{eqnarray}
%%%%%%%%%%%%%%%%%%%%%%%%%%%
where we denote by $X_0=\sqrt{2/d}\mathbb{I}_d$ hereafter. The
elements of $T$ are given by $T_{ij}={\rm Tr}[\mathcal {E}^\dag
(X_i)X_j]/2$ (clearly, $T_{00}=1$ and $T_{0j}=0$ for $j \geq 1$).
This further gives $ x'_i= {\rm Tr} [\rho \mathcal{E}^\dag
(X_i)]=\sum_{j=0}^{d^2-1} T_{ij} x_j$.

% For one-column wide figures use
\begin{figure}
\centering
\resizebox{0.43 \textwidth}{!}{%
\includegraphics{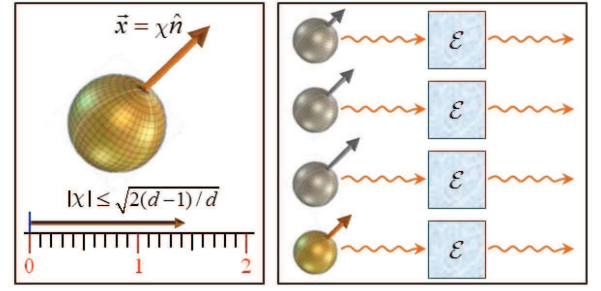}}
% If not, use\vspace{5cm}
% Give the correct figure height in cm
\caption{(Color online) States of the same family
$\{\rho^{\hat{n}}\}$ are represented by the characteristic vectors
$\vec{x}$ along the same or opposite directions (left). When
$\{\rho^{\hat{n}}\}$ traverse a quantum channel $\mathcal {E}$
(right), their decoherence process can be described qualitatively by
that of $\rho_p^{\hat{n}}$ with the unit vector $\hat{n}$ (the
bottommost golden one).} \label{fig:1}
% Give a unique label
\end{figure}
%%%%%%%%%%%%%%%%%%%%%%%%%%%%%%

To present our central result, we first classify the states $\rho$
of Eq. \eqref{eq3-1} into different families $\rho=
\{\rho^{\hat{n}}\}$, with
%%%%%%%%%%%%%%%%%%%%%%%%%%%
\begin{eqnarray}\label{eq3-6}
 \rho^{\hat{n}}=\frac{1}{d}\mathbb{I}_d+\frac{1}{2}\chi\hat{n}\cdot\vec{X},
\end{eqnarray}
%%%%%%%%%%%%%%%%%%%%%%%%%%%
and $\hat{n}=(n_1,n_2,\ldots,n_{d^2-1})$ is a unit vector in
$\mathbb{R}^{d^2-1}$, while $\chi$ is a parameter satisfying $|\chi|
\leq \sqrt{2(d-1)/d}$ as ${\rm Tr}(\rho^{\hat{n}})^2=|\chi|^2/2+
1/d$. By this classification scheme, different families of states
are labeled by different unit vectors $\hat{n}$, while states belong
to the same family $\rho^{\hat{n}}$ are characterized by a common
$\hat{n}$, and can be distinguished by different multiplicative
factors $\chi$ (see Fig. \ref{fig:1}). That is to say,
$\rho^{\hat{n}}$ represents states with the characteristic vectors
$\vec{x}$ along the same or completely opposite directions but
possessing different lengths.

While $\rho^{\hat{n}}$ is fully described by $\chi\hat{n}$,
and the action of $\mathcal {E}$ on $\rho^{\hat{n}}$ can be written
equivalently as the map $\vec{x}'=\mathcal {E}(\chi\hat{n})$, a
quantum property measure $Q$ may only be function of
$\chi\hat{n}^s$, i.e., $Q(\rho^{\hat{n}})= Q(\chi\hat{n}^s)$, with
$\hat{n}^s = \{n_k\}_{k=k_1,\dots,k_\alpha}$ ($\alpha\leq d^2-1$)
the subset of $\hat{n}$. Then as one can always make $Q_{\max}\geq
1$ (otherwise, one can normalize it by simply multiplying a
constant), we have the following lemma.

\textbf{Lemma 1:} For any quantum property measure
$Q(\rho^{\hat{n}})= Q(\chi\hat{n}^s)$ that can be factorized as
$Q(\chi \hat{n}^s)= f(\chi) g(\hat{n}^s)$, and quantum channel
$\mathcal {E}$ that gives the map $\mathcal {E}(\chi
\hat{n}^s)=\chi\mathcal{E}(\hat{n}^s)$, the factorization relation
%%%%%%%%%%%%%%%%%%%%%%%%%%%
\begin{eqnarray}\label{eq3-ad}
 Q[\mathcal {E}(\rho^{\hat{n}})]= Q(\rho^{\hat{n}})
                  Q[\mathcal {E}(\rho_p^{\hat{n}})],
\end{eqnarray}
%%%%%%%%%%%%%%%%%%%%%%%%%%%
holds, where $f(\chi)$ and $g(\hat{n}^s)$ are functionals of $\chi$
and $\hat{n}^s$, respectively, and $\rho_p^{\hat{n}}= \mathbb{I}_d/d
+\chi_p\hat{n} \cdot\vec{X}/2$ is the probe state, with $\chi_p$
solution of the equation $f(\chi_p)g(\hat{n}^s)=1$.

The proof is in the Supplemental Material \cite{supp}. Equipped with
this lemma, we are now in position to present the following theorem.

\textbf{Theorem 1:} If the transformation matrix elements $T_{k0}=0$
for $k \in \{1,2,\ldots,d^2-d\}$, then the evolution of $C_{l_1}
[\mathcal{E} (\rho^{\hat{n}})]$ obeys the following factorization
relation
%%%%%%%%%%%%%%%%%%%%%%%%%%%
\begin{eqnarray}\label{eq3-7}
 C_{l_1}[\mathcal {E}(\rho^{\hat{n}})]= C_{l_1}(\rho^{\hat{n}})
                        C_{l_1}[\mathcal {E}(\rho_p^{\hat{n}})],
\end{eqnarray}
%%%%%%%%%%%%%%%%%%%%%%%%%%%
where the probe state $\rho_p^{\hat{n}}= \mathbb{I}_d/d+
\chi_p\hat{n} \cdot\vec{X}/2$, with $\chi_p = 1/\sum_{r=1}^{d_0}
(n_{2r-1}^2+ n_{2r}^2)^{1/2}$.

The proof to this theorem is also presented in \cite{supp}. Here,
we further demonstrate an equivalent form of it. As $T_{k0} =0$
for $k\in\{1,2,\ldots,d^2-d\}$, we have $\mathcal {E}^\dag (X_k)
= \sum_{j=1}^{d^2-1} T_{kj}X_j$, hence ${\rm Tr} [\mathcal{E}^\dag
(X_k)] =0$. On the other hand, ${\rm Tr}[\mathcal {E}^\dag (X_k)]
={\rm Tr} (X_k \sum_\mu E_\mu E_\mu^\dag)$. This, together with
Eq. \eqref{eq3-2}, require that elements of $A= \sum_\mu E_\mu
E_\mu^\dag$ must satisfy $A_{ij} \pm A_{ji}=0$ for $i \neq j$.
This gives rise to the following corollary:

\textbf{Corollary 1:} If the summation $A=\sum_\mu E_\mu E_\mu^\dag$
is diagonal, i.e., $A = \text{diag} \{A_{11}, A_{22},\ldots,
A_{d,d}\}$, then the evolution of $C_{l_1}[\mathcal{E}
(\rho^{\hat{n}})]$ obeys the factorization relation of Eq.
\eqref{eq3-7}.

This corollary means that in addition to the usual completeness
condition $\sum_\mu E_\mu E_\mu^\dag=\mathbb{I}_d$ of the CPTP map,
the factorization relation \eqref{eq3-7} further requires $\sum_\mu
E_\mu E_\mu^\dag$ to be diagonal. For convenience of later
presentation, we denote this kind of channel $\mathcal
{E}_{\rm F}$, Clearly, it includes the unital channel $\mathcal
{E}_{\rm U}$ [i.e., $\mathcal {E}_{\rm U}(\mathbb{I}_d/d)=
\mathbb{I}_d/d$] as a special case.

From a geometric perspective, Theorem 1 indicates that for all
states of the same family $\rho^{\hat{n}}$, namely, states with the
characteristic vectors $\vec{x}$ along the same or opposite
directions, their decoherence dynamics measured by the $l_1$ norm
can be represented qualitatively by that of the probe state
$\rho_p^{\hat{n}}$, as the magnitude of $C_{l_1}[\mathcal {E}_{\rm
F} (\rho^{\hat{n}})]$ equals the product of the initial coherence
$C_{l_1}(\rho^{\hat{n}})$ and the evolved coherence $C_{l_1}
[\mathcal {E}_{\rm F}(\rho_p^{\hat{n}})]$. This simplifies greatly
the assessment of the decoherence process of an open quantum system.

Moreover, the obtained factorization relation \eqref{eq3-7} provides
a strong link between amount of the coherence loss of an open
system and structures of the applied quantum channels.
Particularly, as $\rho^{\hat{n}}$ with the vectors
$\vec{x}$ along the same or opposite directions fulfill
the same decoherence law, the geometric approach adopted here may
offer a route for better understanding the interplay between
geometry of the state space and various aspects of its quantum
features. It might also provides a deeper insight into the effects
of gate operation in quantum computing and experimental generation
of coherent resources in noisy environments, as $\mathcal {E}_{\rm
F}(\rho^{\hat{n}})$ can specify the actions of environments, of
measurements, or of both on states $\rho^{\hat{n}}$.

It is also worthwhile to mention that when some restrictions are
imposed on the quantum channel, the factorization relation
\eqref{eq3-7} can be further simplified.

\textbf{Corollary 2:} If a channel $\mathcal{E}$ yields $\mathcal{E}
^\dag (X_k)= q(t) X_k$ for $\{X_k\}_{k= k_1, \ldots, k_\beta}$
($\beta \leq d^2-d$), with $q(t)$ containing information on
$\mathcal{E}$'s structure, then the factorization relation
%%%%%%%%%%%%%%%%%%%%%%%%%%%
\begin{equation}\label{eq3-10}
 C_{l_1}[\mathcal {E} (\rho)]=|q(t)| C_{l_1}(\rho),
\end{equation}
%%%%%%%%%%%%%%%%%%%%%%%%%%%
holds for the family of states $\rho = \mathbb{I}_d /d +
\sum_{k=k_1}^{k_\beta} x_k X_k/2+\sum_{l=d^2-d+1}^{d^2-1} x_l
X_l/2$.

The proof of this corollary is direct. As $\mathcal{E}^\dag (X_k)=
q(t) X_k$ for $\{X_k\}_{k= k_1,\ldots, k_\beta}$, the parameters
$x'_k$ for $\mathcal {E}(\rho)$ are given by $x'_k=q(t)x_k$.
Therefore, by Eq. \eqref{eq3-3} we obtain $C_{l_1}[\mathcal {E}
(\rho)]=|q(t) | C_{l_1}(\rho)$. That is to say, the evolution of the
$l_1$ norm of coherence for $\mathcal {E}(\rho)$ is solely
determined by the product of the initial coherence and a noise
parameter $|q(t)|$.

There are many quantum channels satisfying the condition of
Corollary 2. For instance, the Pauli channel $\mathcal {E}_{\rm PL}$
and Gell-Mann channel $\mathcal {E}_{\rm GM}$ given in Ref.
\cite{hufan}, the generalized amplitude damping channel for the
one-qubit states \cite{Nielsen}. Note that $\mathcal {E}_{\rm PL}$
covers the bit flip, phase flip, bit-phase flip, phase damping, and
depolarizing channels which embody typical noisy sources in quantum
information as special cases.

One can also construct quantum channel $\mathcal {E}_{\rm G}$ under
the action of which $C_{l_1}[\mathcal {E}_{\rm G}(\rho)]$ obeys the
factorization relation \eqref{eq3-10} for arbitrary initial state.
The Kraus operators describing
$\mathcal{E}_{\rm G}$ are given by
%%%%%%%%%%%%%%%%%%%%%%%%%%%
\begin{equation}\label{eq3-11}
 \begin{split}
  & E_0^{\rm G} = \frac{1}{d}\sqrt{1 + (d^2-d) q + (d-1)q_0}\mathbb{I}_d, \\
  & E_k^{\rm G} = \sqrt{\frac{1 - q_0}{2d}}X_a,\\
  & E_l^{\rm G} = \sqrt{\frac{1 - d q + (d-1)q_0}{2d}}X_b,
 \end{split}
\end{equation}
%%%%%%%%%%%%%%%%%%%%%%%%%%%
with $k\in\{1,\ldots,d^2-d\}$, and $l\in\{d^2-d+1,\ldots,d^2-1\}$,
while $q(t)$ and $q_0(t)$ are the time-dependent noisy parameters.
Clearly, $\mathcal {E}_{\rm G}$ includes the depolarizing channel
(i.e., $q_0=q$) as a special case.

\emph{N-qubit case.}--- A general $N$-qubit state $\rho$ can be
written as $\rho = \mathbb{I}_{2^N} /2^N+ \vec{y}\cdot\vec{Y}/2$,
where $\vec{Y}=\{Y_1,Y_2,\dots,Y_{4^N-1}\}$, and
%%%%%%%%%%%%%%%%%%%%%%%%%%%
\begin{equation}\label{eq3-12}
 Y_j=2^{(1-N)/2}\sigma_{j_1}\otimes\sigma_{j_2}\otimes\ldots\otimes\sigma_{j_N},
\end{equation}
%%%%%%%%%%%%%%%%%%%%%%%%%%%
with $(j_1 j_2 \ldots j_N)$ taking the values of $(00 \ldots 1)$,
$(00\ldots 2)$, $(00\ldots 3)$, $\ldots$, $(33\ldots 3)$.

We now show that every family of the $N$-qubit states
$\rho^{\hat{m}}=\mathbb{I}_{2^N} /2^N+\chi\hat{m} \cdot\vec{Y}/2$
[$\hat{m}=(m_1, m_2,\ldots,m_{4^N-1})$ is a unit vector] can be
generated by an auxiliary channel $\mathcal{E}_{\rm aux}$ acting on
$\rho$. To this end, we consider $\mathcal{E}_{\rm aux}$ described
by the Kraus operators $E_\mu^{\rm aux}=\sqrt{\varepsilon_\mu}
Y_\mu$, with $\mu\in\{0,1,\ldots, 4^N-1\}$ and $Y_0=\sqrt{2^{1-N}}
\mathbb{I}_{2^N}$. Then, by employing the anticommutation relation
of the Pauli operators, we obtain
%%%%%%%%%%%%%%%%%%%%%%%%%%%
\begin{equation}\label{eq3-13}
\mathcal{E}_{\rm aus}^\dag(Y_\nu)=\sum_{\mu=0}^{4^N-1} c_{\nu\mu}
\varepsilon_\mu Y_\nu,
\end{equation}
%%%%%%%%%%%%%%%%%%%%%%%%%%%
where $c_{\nu\mu}= 2^{1-N}(-1)^{\sum_{k=1}^{N}\xi^{\nu\mu}_k}$, with
$\xi^{\nu\mu}_k= 0$ if $\nu_k \mu_k (\nu_k-\mu_k)= 0$, and
$\xi^{\nu\mu}_k= 1$ otherwise. This formula is equivalent to
$\mathcal {E}_{\rm aux}^\dag(Y_\nu)=q_\nu Y_\nu$, with $q_\nu=
\sum_{\mu} c_{\nu\mu} \varepsilon_\mu $ encoding information of
$\mathcal {E}_{\rm aux}$. To solve $\{\varepsilon_\mu\}$, we
introduce the coefficient matrix $\hat{c}=\{c_{\nu\mu}\}$, the
column vectors $\hat{\varepsilon}= (\varepsilon_0,\varepsilon_1,
\ldots, \varepsilon_{4^N-1})^T$, and $\hat{q}= (1,q_1,\ldots,
q_{4^N-1})^T$, then $\hat{c}\hat{\varepsilon}= \hat{q}$, and
$\varepsilon$ can be derived as $\hat{\varepsilon}=\hat{c}^{-1}
\hat{q}$. By choosing $q_\nu=\chi m_\nu/y_\nu$, we obtain $y'_\nu=
\chi m_\nu$. Therefore, we have the third corollary.

\textbf{Corollary 3:} For any $N$-qubit state $\rho$, one can
construct a quantum channel $\mathcal {E}_{\rm aux}$ such that
$\rho^{\hat{m}}= \mathcal {E}_{\rm aux}(\rho)$.

This corollary, together with Eq. \eqref{eq3-7}, implies that
%%%%%%%%%%%%%%%%%%%%%%%%%%%
\begin{eqnarray}\label{eq3-14}
 C_{l_1}[\mathcal{E}_{\rm F}\mathcal{E}_{\rm aux}(\rho)]= C_{l_1}
        [\mathcal{E}_{\rm aux}(\rho)]C_{l_1}[\mathcal{E}_{\rm F}(\rho_p^{\hat{m}})],
\end{eqnarray}
%%%%%%%%%%%%%%%%%%%%%%%%%%%
with $\rho_p^{\hat{m}}=\mathbb{I}_{2^N}
/2^N+\chi_p\hat{m}\cdot\vec{Y}/2$, $\chi_p= 1/\sum_{r=1}^{d_0}
(\bar{m}_{2r-1}^2+\bar{m}_{2r}^2) ^{1/2}$, $d_0=(4^N-2^N)/2$, and
$\bar{m}_i=\sum_j a_{ij}m_j$, with $a_{ij}$ being determined by the
transformation between $\{Y_j\}$ and $\{X_i\}$: $X_i=\sum_j a_{ij}
Y_j$. This equality generalizes the result of Eq. \eqref{eq3-7} for
the $N$-qubit states $\rho$. It shows that the coherence of the
evolved state under the actions of two cascaded channels $\mathcal
{E}_{\rm F}\mathcal {E}_{\rm aux}$ is determined by the product of
the coherence for the evolved probe state under the action of
$\mathcal {E}_{\rm F}$ and the coherence for the generated state by
$\mathcal {E}_{\rm aux}$.

As every $Y_j$ in Eq. \eqref{eq3-12} can always be decomposed as
linear combinations of the generators $\{X_i\}$, the above result
applies also to the qudit states with $d=2^N$. As an explicit
example, the transformation between $\{Y_j\}$ and $\{X_i\}$ for
$N=2$ is given in the Supplemental Material \cite{supp}, from which
$\mathcal {E}_{\rm aux}$ and $\{a_{ij}\}$ can be constructed
directly.

\emph{Frozen coherence.}--- By Theorem 1 we can also derive
conditions on the quantum channel for which the $l_1$ norm of
coherence is frozen. To elucidate this, we return to Eq.
\eqref{eq3-7}, from which one can see that $C_{l_1}[\mathcal {E}
(\rho^{\hat{n}})]$ is frozen if the coherence of the probe state
remains constant 1 during the evolution, i.e., $C_{l_1}
[\mathcal {E}(\rho_p^{\hat{n}})]\equiv 1$. For later use, we
denote by $T^S$ the submatrix of $T$ consisting $T_{ij}$ with $i$
ranging from 1 to $d^2-d$ and $j$ from 1 to $d^2-1$. Then by Theorem
1 and the reasoning in its proof \cite{supp}, we obtain the fourth
corollary.

\textbf{Corollary 4:} If $T_{k0}=0$ for $k\in \{1,2,\ldots,d^2-d\}$,
and $T^S$ is a rectangular block diagonal matrix, with the main
diagonal blocks
%%%%%%%%%%%%%%%%%%%%%%%%%%%
\begin{equation}\label{eq3-16}
 T^S_r =\left(\begin{array}{cc}
              T_{2r-1,2r-1}  &   T_{2r-1,2r} \\
              T_{2r,2r-1}    &   T_{2r,2r}
              \end{array}\right)
              ~(r\in \{1,\dots,d_0\}),
\end{equation}
%%%%%%%%%%%%%%%%%%%%%%%%%%%
being orthogonal matrices, i.e., $(T^S_r)^T T^S_r = \mathbb{I}_2$,
the $l_1$ norm of coherence for $\rho^{\hat{n}}$ will be frozen
during the entire evolution.

The proof of this corollary is in
\cite{supp}. It enables one to construct quantum channels for which
the $l_1$ norm of coherence is frozen. As an explicit example, we
consider the one-qubit state case, with $\mathcal {E}$ being
described by $E_i= \sum_{j=0}^3 \varepsilon_{ij}
\sigma_j$, $i\in\{0,1,2,3\}$ and $\varepsilon _{ij}\in \mathbb{C}$.
Then by Corollary 4, one can obtain that when
$\varepsilon_{i0}= \varepsilon_{i3}=0$, and $\sum_i
|\varepsilon_{i1}+i\varepsilon_{i2}| ^2= \sum_i |\varepsilon_{i1}-
i\varepsilon_{i2}|^2=[\sum_i (|\varepsilon_{i1}|^2-
|\varepsilon_{i2}|^2)]^2+ 4 [\sum_i {\rm Re} (\varepsilon_{i1}^*
\varepsilon_{i2})]^2 =1$, or when $\varepsilon_{i1}=\varepsilon_{i2}
=0$, and $\sum_i |\varepsilon_{i0}+ \varepsilon_{i3}|^2=\sum_i
|\varepsilon_{i0} -\varepsilon_{i3}|^2 = [\sum_i
(|\varepsilon_{i0}|^2- |\varepsilon_{i3}|^2)]^2 +4[\sum_i {\rm
Im}(\varepsilon_{i3}^* \varepsilon_{i0})]^2=1$, with ${\rm
Re}(\cdot)$ and ${\rm Im}(\cdot)$ representing, respectively, the
real and imaginary parts of a number, the $l_1$ norm of coherence
will be frozen \cite{supp}. There are a host of
$\{\varepsilon_{ij}\}$ that fulfill the requirements, e.g., the
simplest case of $\varepsilon_{01}= q(t)$, $\varepsilon_{02}=\pm
q'(t)$, $\varepsilon_{i1}= \varepsilon_{i2}=0$, or $\varepsilon_{00}=q(t)$,
$\varepsilon_{03}=\pm i q'(t)$, $\varepsilon_{i0}= \varepsilon_{i3}=0$,
with $i\in\{1,2,3\}$, $q'(t)=\sqrt{1-q^2(t)}$, and $q(t)$ contains the
information on $\mathcal {E}$'s structure and its coupling with the system.

Moreover, for certain special initial states, the freezing condition
presented in Corollary 4 may be further relaxed. In fact, for
$\rho^{\hat{n}}$ with certain $n_{2r-1} =0$ (or $n_{2r}=0$), the
freezing condition $(T^S_r)^T T^S_r = \mathbb{I}_2$ simplifies to
$T_{2r-1,2r}^2+ T_{2r,2r}^2 =1$ (or $T_{2r-1,2r-1}^2+ T_{2r, 2r-1}^2
=1$) \cite{supp}. For instance, when considering the channel
$\mathcal{E}_{\rm PL}$ \cite{hufan}, the $l_1$ norm of coherence for
$\rho^{\hat{n}}$ with $n_2= 0$ is frozen during the entire evolution
when $q_1=1$ (i.e., the bit flip channel). Similarly, for
$\rho^{\hat{n}}$ with $n_1= 0$, it is frozen when $q_2=1$ (i.e., the
bit-phase flip channel). These are in facts the results obtained in
Ref. \cite{frozen}. Needless to say, when $(T^S_r)^T T^S_r =
\mathbb{I}_2$, the $l_1$ norm of coherence is also frozen for
$\rho^{\hat{n}}$ with certain $n_{2r-1} =0$ or $n_{2r} =0$.

Finally, we remark that the coherence concurrence $C_z(\rho)$ which
is a monotonic function of the intrinsic randomness coherence
measure for the one-qubit states \cite{meas4}, and the trace norm of
coherence $C_{\rm tr}(\rho)$ for certain special sates
\cite{meas3,frozen}, coincide with the $l_1$ norm of coherence.
Hence, our results presented in this Letter also apply to them.
Moreover, the $l_1$ norm of coherence is intimately related to the
negativity of quantumness \cite{nega,frozen}, and is a monotone of
the entanglement-based coherence measure $C_g(\rho)$ for $\rho$ of
one qubit \cite{meas2}. For these limited cases, our results also
provide a route for inspecting interplay between peculiar decay
behaviors of coherence, quantumness, and entanglement.

Apart from quantum coherence, there are other quantifiers fulfilling
conditions of Lemma 1, hence the factorization relation \eqref{eq3-ad}
holds. Some examples encompass the purity monotone, the geometric
quantum discord \cite{gqd}, the measurement-induced nonlocality
\cite{min}, the Hellinger distance discord \cite{square}, the
maximum Bell-inequality violation \cite{Bell}, and fidelity of remote state
preparation \cite{rsp} and quantum teleportation \cite{qt} (see \cite {supp}
for more detail). These manifest again the universality of the
factorization relation obtained in this Letter, and will certainly
deepen our understanding of the already rich and appealing subject
of quantum channels or the CPTP maps.

\emph{Summary.}--- We have established a simple factorization
relation for the evolution equation of the $l_1$ norm of coherence.
This relation is of practical relevance for assessing coherence loss
of an open quantum system. For a general $d$-dimensional state, we
determined condition such that the factorization relation holds.
This condition can be described as a restriction on the form of the
transformation matrix, or on the summation of the product $E_\mu
E_\mu^\dag$ of the Kraus operators, of the quantum channel. By
introducing an auxiliary channel, we further presented a more
general relation which applies to all the $N$-qubit states.
Moreover, we have also determined a condition the transformation
matrix should satisfy such that the $l_1$ norm of coherence for a
general $d$-dimensional state is dynamically frozen, and constructed
explicitly the desired quantum channels for states of one qubit. We
hope these results may help in understanding the interplay between
structure of the quantum channel, geometry of the state space, and
decoherence of an open system, as well as their combined effects on
peculiar decay behaviors of various quantum correlations.

This work was supported by NSFC (Grants No. 11205121, 91536108), NSF
of Shaanxi Province (Grant No. 2014JM1008), the Youth Foundation of
XUPT, and CAS (Grant No. XDB01010000).

\newcommand{\PRL}{Phys. Rev. Lett. }
\newcommand{\RMP}{Rev. Mod. Phys. }
\newcommand{\PRA}{Phys. Rev. A }
\newcommand{\PRB}{Phys. Rev. B }
\newcommand{\PRE}{Phys. Rev. E }
\newcommand{\PRX}{Phys. Rev. X }
\newcommand{\NJP}{New J. Phys. }
\newcommand{\JPA}{J. Phys. A }
\newcommand{\JPB}{J. Phys. B }
\newcommand{\PLA}{Phys. Lett. A }
\newcommand{\NP}{Nat. Phys. }
\newcommand{\NC}{Nat. Commun. }
%

% BibTeX users please use
% \bibliographystyle{}
% \bibliography{}
% Non-BibTeX users please use

\newpage
\setcounter{equation}{0}
\renewcommand{\theequation}{A\arabic{equation}}
\begin{appendix}

\section{Supplemental Material}

\subsection{Proof of Lemma 1}
\emph{Proof.} As the channel $\mathcal {E}$ gives the map
%%%%%%%%%%%%%%%%%%%%%%%%%%%
\begin{equation}\label{eq-s1}
 \mathcal {E}(\chi \hat{n}^s)=\chi\mathcal{E}(\hat{n}^s),
\end{equation}
%%%%%%%%%%%%%%%%%%%%%%%%%%%
for $\chi\in \mathbb{R}$ and $\hat{n}^s = \{n_k\}_{k=k_1,\dots,
k_\alpha}$ ($\alpha\leq d^2-1$) , and the measure $Q(\rho^{\hat{n}})
=Q(\chi\hat{n}^s)$ fulfills
%%%%%%%%%%%%%%%%%%%%%%%%%%%
\begin{equation}\label{eq-sp1}
 Q(\chi\hat{n}^s)=f(\chi) g(\hat{n}^s),
\end{equation}
%%%%%%%%%%%%%%%%%%%%%%%%%%%
we have
%%%%%%%%%%%%%%%%%%%%%%%%%%%
\begin{equation}\label{eq-s2}
  \begin{split}
   & Q[\mathcal{E}(\rho^{\hat{n}})]= Q[\chi\mathcal{E}(\hat{n}^{s})]
                                   = f(\chi)g[\mathcal{E}(\hat{n}^{s})],\\
   & Q[\mathcal{E}(\rho_p^{\hat{n}})]= Q[\chi_p \mathcal{E}(\hat{n}^{s})]
                                     = f(\chi_p)g[\mathcal{E}(\hat{n}^{s})].
  \end{split}
\end{equation}
%%%%%%%%%%%%%%%%%%%%%%%%%%%
Hence, it is evident that
%%%%%%%%%%%%%%%%%%%%%%%%%%%
\begin{equation}\label{eq-s3}
 Q[\mathcal {E}(\rho^{\hat{n}})]=Q(\rho^{\hat{n}})Q[\mathcal{E}
                                 (\rho_p^{\hat{n}})],
\end{equation}
%%%%%%%%%%%%%%%%%%%%%%%%%%%
when $f(\chi_p)g(\hat{n}^s)=1$ with respect to $\chi_p$ is solvable.

If the maximum $Q_{\rm max} \geqslant 1$, the equation $f(\chi_p)
g(\hat{n}^s)=1$ is always solvable as $f(\chi_p)g(\hat{n}^s)=
Q(\rho_p^{\hat{n}})$. If $Q_{\rm max}<1$, one can normalize it by
simply introducing a constant $N$ such that $Q'_{\rm max}=N Q_{\rm
max}=1$. Now, $Q'$ obeys the factorization relation of Eq. (8) in
the main text. \hfill{$\square$}

\subsection{Proof of Theorem 1}
\emph{Proof.} First, by using Eq. (4) of the main text and the fact
that $\vec{x} = \chi \hat{n}$, we obtain
%%%%%%%%%%%%%%%%%%%%%%%%%%%
\begin{equation}\label{eq-s4}
 C_{l_1}(\rho^{\hat{n}})= \chi \sum_{r=1}^{d_0} \sqrt{n_{2r-1}^2+n_{2r}^2}.
\end{equation}
%%%%%%%%%%%%%%%%%%%%%%%%%%%
which corresponds to $C_{l_1}(\rho^{\hat{n}})=f(\chi)g(\hat{n}^s)$,
with $f(\chi)=\chi$ and $g(\hat{n}^s)= \sum_{r=1}^{d_0}
(n_{2r-1}^2+n_{2r}^2)^{1/2}$.

Second, when the transformation matrix elements $T_{k0}=0$ for $k
\in \{1,2,\ldots,d^2-d\}$, we have
%%%%%%%%%%%%%%%%%%%%%%%%%%%
\begin{equation}\label{eq-sp4}
 x'_k=\chi\sum_{j=1}^{d^2-1}T_{kj}n_j=\chi\mathcal {E}(n_k),
\end{equation}
%%%%%%%%%%%%%%%%%%%%%%%%%%%
and therefore $\mathcal {E}(\chi \hat{n}^s)=\chi\mathcal{E}
(\hat{n}^s)$.

From Eqs. \eqref{eq-s4} and \eqref{eq-sp4} one can see that both the
$l_1$ norm of coherence and the quantum channel $\mathcal {E}$
fulfill the requirements of Lemma 1. Thus, the factorization
relation (9) of the main text holds.

Moreover, the probe state $\rho_p^{\hat{n}}= \mathbb{I}_d/d+\chi_p
\hat{n} \cdot\vec{X}/2$, with $\chi_p$ being solution of the
equation
%%%%%%%%%%%%%%%%%%%%%%%%%%%
\begin{equation}\label{eq-s5}
 \chi_p g(\hat{n}^s)=1,
\end{equation}
%%%%%%%%%%%%%%%%%%%%%%%%%%%
which can be solved as $\chi_p  = 1/\sum_{r=1}^{d_0}(n_{2r-1}^2
+n_{2r}^2)^{1/2}$. This completes the proof. \hfill{$\square $}

\subsection{Transformation between $\{Y_j\}$ and $\{X_i\}$}
We list here the transformation between generators $\{Y_j\}$ for the
two-qubit states and $\{X_i\}$ for the qudit states with $d=4$. They
are as follows:
%%%%%%%%%%%%%%%%%%%%%%%%%%%
\begin{equation}\label{eq-s6}
 \begin{split}
  & Y_{1,13}=\frac{1}{\sqrt{2}}(X_1 \pm X_{11}),\\
  & Y_{2,14}=\frac{1}{\sqrt{2}}(X_2 \pm X_{12}),\\
  & Y_{4,7}=\frac{1}{\sqrt{2}}(X_3 \pm X_9), \\
  & Y_{5,10}=\frac{1}{\sqrt{2}}(X_7 \pm X_5),\\
  & Y_{9,6}=\frac{1}{\sqrt{2}}(X_6 \pm X_{8}), \\
  & Y_{8,11}=\frac{1}{\sqrt{2}}(X_4 \pm X_{10}),\\
  & Y_{12}=\frac{\sqrt{6}}{3}X_{14}+\frac{1}{\sqrt{3}}X_{15},\\
  & Y_{3,15}=\frac{1}{\sqrt{2}}X_{13}\mp \frac{1}{\sqrt{6}}
             X_{14}\pm\frac{1}{\sqrt{3}}X_{15},
 \end{split}
\end{equation}
%%%%%%%%%%%%%%%%%%%%%%%%%%%
where we have arranged elements $X_i$ of $\vec{X}$ in the sequence
of $\vec{X}=\{u_{12},v_{12}, u_{13},v_{13}, \ldots, u_{34}, v_{34},
w_1,w_2,w_3\}$, and elements $Y_j=2^{-1/2}\sigma_{j_1}\otimes
\sigma_{j_2}$ of $\vec{Y}$ with $(j_1 j_2)$ in the sequence of
$(01)$, $(02)$, $(03)$, $(10)$, $(11)$,  $(12)$, $(13)$, $\ldots$,
$(33)$.

\subsection{Proof of Corollary 4}
\emph{Proof.} First, as the submatrix $T^S$ is rectangular block
diagonal, the elements $T_{ij}$ in the off-diagonal blocks are all
zero. This, together with $T_{k0}=0$ for $k\in \{1,2, \ldots,
d^2-d\}$, gives rise to
%%%%%%%%%%%%%%%%%%%%%%%%%%%
\begin{equation}\label{eq-s7}
 \begin{split}
   n'_{2r-1} &=\sum_{j=0}^{d^2-1}T_{2r-1,j} n_j \\
             &= T_{2r-1,2r-1}n_{2r-1}+T_{2r-1,2r}n_{2r}, \\
     n'_{2r} &= \sum_{j=0}^{d^2-1} T_{2r,j} n_j \\
             &= T_{2r,2r-1}n_{2r-1}+T_{2r,2r}n_{2r},
 \end{split}
\end{equation}
%%%%%%%%%%%%%%%%%%%%%%%%%%%
for $r\in \{1,2, \ldots, d_0\}$.

Second, the requirement that any $2\times 2$ block $T^S_r$ is an
orthogonal matrix, i.e., $(T^S_r)^T T^S_r = \mathbb{I}_2$, yields
%%%%%%%%%%%%%%%%%%%%%%%%%%%
\begin{equation}\label{eq-s8}
 \begin{split}
  & T_{2r-1,2r-1}^2+ T_{2r,2r-1}^2 = T_{2r-1,2r}^2+ T_{2r,2r}^2 =1,\\
  & T_{2r-1,2r-1}T_{2r-1,2r}+T_{2r,2r-1}T_{2r,2r}=0.
 \end{split}
\end{equation}
%%%%%%%%%%%%%%%%%%%%%%%%%%%

By combining the above two equations, it is straightforward to
observe that
%%%%%%%%%%%%%%%%%%%%%%%%%%%
\begin{equation}\label{eq-s9}
 \begin{split}
 n_{2r-1}'^2+n_{2r}'^2 & = (T_{2r-1,2r-1}^2+T_{2r,2r-1}^2)n_{2r-1}^2 \\
                       &~~~~~  + (T_{2r-1,2r}^2+ T_{2r,2r}^2)n_{2r}^2 \\
                       &~~~~~  + 2(T_{2r-1,2r-1}T_{2r-1,2r} \\
                       &~~~~~  + T_{2r,2r-1}T_{2r,2r})n_{2r-1}n_{2r} \\
                       & = n_{2r-1}^2+n_{2r}^2,
 \end{split}
\end{equation}
%%%%%%%%%%%%%%%%%%%%%%%%%%%
and therefore from Eq. \eqref{eq-s4} we have $C_{l_1}[\mathcal
{E}(\rho_p^{\hat{n}})]\equiv 1$. This, together with Theorem 1,
implies
%%%%%%%%%%%%%%%%%%%%%%%%%%%
\begin{equation}\label{eq-s10}
 C_{l_1}[\mathcal{E} (\rho^{\hat{n}})] = C_{l_1}(\rho^{\hat{n}})
\end{equation}
%%%%%%%%%%%%%%%%%%%%%%%%%%%
and hence completes the proof. \hfill{$\square $}

From the reasoning in the above proof, it is also worthwhile to note
that for the initial quantum states with certain $n_{2r-1} =0$ (or
$n_{2r} =0$), we have $n_{2r-1}'^2+n_{2r}'^2=n_{2r-1}^2+n_{2r}^2$
when $T_{2r-1,2r}^2+ T_{2r,2r}^2=1$ (or $T_{2r-1,2r-1}^2+
T_{2r,2r-1}^2 =1$), hence the requirement $(T^S_r)^T T^S_r =
\mathbb{I}_2$ is relaxed compared with that of the general initial
states. Needless to say, when $(T^S_r)^T T^S_r = \mathbb{I}_2$, the
coherence is also frozen for states $\rho^{\hat{n}}$ with certain
$n_{2r-1} =0$ or $n_{2r} =0$.

\subsection{Frozen quantum coherence of one qubit}
We set out to construct quantum channel $\mathcal {E}$ under the
action of which the $l_1$ norm of coherence is frozen during the
entire evolution. To this end, we let
%%%%%%%%%%%%%%%%%%%%%%%%%%%
\begin{equation}\label{eq-p1}
 E_i= \sum_{j=0}^3 \varepsilon_{ij} \sigma_j ~(i\in\{0,1,2,3\}),
\end{equation}
%%%%%%%%%%%%%%%%%%%%%%%%%%%
be the Kraus operators of $\mathcal {E}$, where $\varepsilon_{ij}
\in \mathbb{C}$, and their values should satisfy certain constraints
such that the requirement of Corollary 4 is satisfied.

First, the completeness condition of the CPTP map, namely, $\sum_i
E_i^\dag E_i =\mathbb{I}$ \cite{Nielsen-sp}, requires
%%%%%%%%%%%%%%%%%%%%%%%%%%%
\begin{equation}\label{eq-p2}
 \begin{split}
  & \sum_i (|\varepsilon_{i0}+\varepsilon_{i3}|^2+|\varepsilon_{i1}+i\epsilon_{i2}|^2)=1, \\
  & \sum_i (|\varepsilon_{i0}-\varepsilon_{i3}|^2+|\varepsilon_{i1}-i\varepsilon_{i2}|^2)=1, \\
  & \sum_i [{\rm Re}(\varepsilon_{i0}^*\varepsilon_{i1})+{\rm Im}(\varepsilon_{i3}^*
            \varepsilon_{i2})-i{\rm Re}(\varepsilon_{i0}^*\varepsilon_{i2}) \\
  &~~~~~~~ +i{\rm Im}(\varepsilon_{i3}^*\varepsilon_{i1})]=0,
 \end{split}
\end{equation}
%%%%%%%%%%%%%%%%%%%%%%%%%%%
where $\varepsilon_{ij}^*$ represents conjugation of
$\varepsilon_{ij}$, while ${\rm Re}(\cdot)$ and ${\rm Im}(\cdot)$
represent, respectively, the real and imaginary parts of a number,
and the notation $i$ before $\varepsilon_{i2}$, ${\rm Re}(\cdot)$,
and ${\rm Im}(\cdot)$ is the imaginary unit.

Second, Corollary 4 requires $T_{10}=T_{20}=0$, and $T^S$ to be a
rectangular block diagonal matrix which corresponds to $T_{13}=
T_{23}=0$. This yields
%%%%%%%%%%%%%%%%%%%%%%%%%%%
\begin{equation}\label{eq-p3}
 \begin{split}
  & \sum_i E_i^\dag \sigma_1 E_i =T_{11}\sigma_1+T_{12}\sigma_2, \\
  & \sum_i E_i^\dag \sigma_2 E_i =T_{21}\sigma_1+T_{22}\sigma_2,
\end{split}
\end{equation}
%%%%%%%%%%%%%%%%%%%%%%%%%%%
from which one can obtain
%%%%%%%%%%%%%%%%%%%%%%%%%%%
\begin{equation}\label{eq-p4}
 \begin{split}
  & \sum_i {\rm Re} [(\varepsilon_{i0}+\varepsilon_{i3})(\varepsilon_{i1}^*-i\varepsilon_{i2}^*)]=0, \\
  & \sum_i {\rm Re} [(\varepsilon_{i0}^*-\varepsilon_{i3}^*)(\varepsilon_{i1}-i\varepsilon_{i2})]=0, \\
  & \sum_i {\rm Im} [(\varepsilon_{i0}+\varepsilon_{i3})(\varepsilon_{i1}^*-i\varepsilon_{i2}^*)]=0, \\
  & \sum_i {\rm Im} [(\varepsilon_{i0}^*-\varepsilon_{i3}^*)(\varepsilon_{i1}-i\varepsilon_{i2})]=0,
\end{split}
\end{equation}
%%%%%%%%%%%%%%%%%%%%%%%%%%%
and
%%%%%%%%%%%%%%%%%%%%%%%%%%%
\begin{equation}\label{eq-p5}
 \begin{split}
  & T_{11} =\sum_i [|\varepsilon_{i0}|^2+|\varepsilon_{i1}|^2-|\varepsilon_{i2}|^2-|\varepsilon_{i3}|^2], \\
  & T_{22} =\sum_i [|\varepsilon_{i0}|^2-|\varepsilon_{i1}|^2+|\varepsilon_{i2}|^2-|\varepsilon_{i3}|^2], \\
  & T_{12} =2\sum_i[{\rm Re}(\varepsilon_{i1}^*\varepsilon_{i2})-{\rm Im}(\varepsilon_{i3}^*\varepsilon_{i0})], \\
  & T_{21} =2\sum_i[{\rm Re}(\varepsilon_{i1}^*\varepsilon_{i2})+{\rm Im}(\varepsilon_{i3}^*\varepsilon_{i0})]. \\
\end{split}
\end{equation}
%%%%%%%%%%%%%%%%%%%%%%%%%%%

By comparing Eqs. \eqref{eq-p2} and \eqref{eq-p4}, one can note that
the equalities are satisfied when $\varepsilon_{i0}=\varepsilon_{i3}
=0$, $\sum_i |\varepsilon_{i1}+i\varepsilon_{i2}|^2=\sum_i
|\varepsilon_{i1}- i\varepsilon_{i2}|^2=1$, or when
$\varepsilon_{i1}= \varepsilon_{i2} =0$, $\sum_i |\varepsilon_{i0}+
\varepsilon_{i3}|^2 =\sum_i |\varepsilon_{i0}
-\varepsilon_{i3}|^2=1$. Under these two constraints, Eq.
\eqref{eq-p5} simplifies, respectively, to
%%%%%%%%%%%%%%%%%%%%%%%%%%%
\begin{equation}\label{eq-p8}
 \begin{split}
  & T_{11} = -T_{22}=\sum_i (|\varepsilon_{i1}|^2-|\varepsilon_{i2}|^2), \\
  & T_{12} = T_{21} =2\sum_i {\rm Re}(\varepsilon_{i1}^*\varepsilon_{i2}), \\
\end{split}
\end{equation}
%%%%%%%%%%%%%%%%%%%%%%%%%%%
and
%%%%%%%%%%%%%%%%%%%%%%%%%%%
\begin{equation}\label{eq-p9}
 \begin{split}
  & T_{11} = T_{22}=\sum_i (|\varepsilon_{i0}|^2-|\varepsilon_{i3}|^2), \\
  & T_{12} = -T_{21} =-2\sum_i {\rm Im}(\varepsilon_{i3}^*\varepsilon_{i0}). \\
\end{split}
\end{equation}
%%%%%%%%%%%%%%%%%%%%%%%%%%%

Finally, the requirement that $T^S_r$ should be an orthogonal
matrix, i.e., $(T^S_r)^T T^S_r= \mathbb{I}_2$, corresponds to
%%%%%%%%%%%%%%%%%%%%%%%%%%%
\begin{equation}\label{eq-p10}
 \begin{split}
  & |T_{11}|^2+ |T_{21}|^2 =1, \\
  & |T_{12}|^2+ |T_{22}|^2 =1, \\
  & T_{11}^*T_{12}+T_{21}^*T_{22}=0,
\end{split}
\end{equation}
%%%%%%%%%%%%%%%%%%%%%%%%%%%
then from Eqs. \eqref{eq-p8} and \eqref{eq-p9}, one can see that the
equality in the third line of Eq. \eqref{eq-p10} is always
satisfied, while the equalities in the first two lines are
equivalent. Therefore, to freeze the $l_1$ norm of quantum
coherence, the parameters $\varepsilon_{ij}$ should satisfy one of
the following two conditions:
%%%%%%%%%%%%%%%%%%%%%%%%%%%
\begin{equation}\label{eq-p11}
 \begin{split}
  & \varepsilon_{i0}=\varepsilon_{i3} =0~~ ({\rm for}~i\in\{0,1,2,3\}), \\
  & \sum_i |\varepsilon_{i1}+i\varepsilon_{i2}|^2=\sum_i |\varepsilon_{i1}-i\varepsilon_{i2}|^2=1, \\
  & \left[\sum_i (|\varepsilon_{i1}|^2-|\varepsilon_{i2}|^2)\right]^2 +4\left[\sum_i
    {\rm Re}(\varepsilon_{i1}^*\varepsilon_{i2})\right]^2=1,
\end{split}
\end{equation}
%%%%%%%%%%%%%%%%%%%%%%%%%%%
or
%%%%%%%%%%%%%%%%%%%%%%%%%%%
\begin{equation}\label{eq-p12}
 \begin{split}
  & \varepsilon_{i1}=\varepsilon_{i2} =0~ (\text{for}~ i\in\{0,1,2,3\}), \\
  & \sum_i |\varepsilon_{i0}+\varepsilon_{i3}|^2=\sum_i |\varepsilon_{i0}-\epsilon_{i3}|^2=1,\\
  & \left[\sum_i (|\varepsilon_{i0}|^2-|\varepsilon_{i3}|^2)\right]^2 +4\left[\sum_i
    {\rm Im}(\varepsilon_{i3}^*\varepsilon_{i0})\right]^2=1
\end{split}
\end{equation}
%%%%%%%%%%%%%%%%%%%%%%%%%%%

By Eqs. \eqref{eq-p11} and \eqref{eq-p12}, one can construct a host
of quantum channels $\mathcal {E}$ under the action of which the
$l_1$ norm of coherence for the one-qubit states is frozen. For
instance, when $\varepsilon_{01}=q(t)$, $\varepsilon_{02}=\pm
\sqrt{1-q^2(t)}$, and $\varepsilon_{i1}= \varepsilon_{i2}=0$
($i\in\{1,2,3\}$), or when $\varepsilon_{00}=q(t)$,
$\varepsilon_{03}=\pm i\sqrt{1-q^2(t)}$, and $\varepsilon_{i0}=
\varepsilon_{i3}=0$ ($i\in\{1,2,3\}$), we always have
$C_{l_1}[\mathcal {E}(\rho^{\hat{n}})]= C_{l_1} (\rho^{\hat{n}})$.
Here, $q(t)$ is a time-dependent parameter containing information on
$\mathcal {E}$ and its coupling with the system.

\subsection{Other measures fulfilling the factorization relation}
Apart from the $l_1$ norm of coherence, another quantifier
fulfilling the requirement of Lemma 1 is $\mathcal {P}(\rho)$ which
is a monotonic function of the purity $P(\rho)={\rm Tr}\rho^2$ of a
state
%%%%%%%%%%%%%%%%%%%%%%%%%%%
\begin{equation}\label{eq-m1}
 \mathcal {P}(\rho)=P(\rho)-\frac{1}{d}=\frac{1}{2}|\vec{x}|^2.
\end{equation}
%%%%%%%%%%%%%%%%%%%%%%%%%%%
In analogy to Eq. (8), here we have
%%%%%%%%%%%%%%%%%%%%%%%%%%%
\begin{equation}\label{eq-m2}
 \mathcal{P}[\mathcal{E}(\rho)]=\mathcal{P}(\rho)\mathcal{P}[\mathcal{E}(\rho_{p})],
\end{equation}
%%%%%%%%%%%%%%%%%%%%%%%%%%%
with $\rho_{p}$ bing the probe state for which $|\vec{x}_p|=
\sqrt{2}$.

Even for bipartite states, there are quantum property measures
fulfilling the requirement of Lemma 1, hence the factorization
relation (8) holds for them. Examples of this kind of measures
encompass the discord-like correlations, such as the geometric
quantum discord defined via the Schatten 2-norm (i.e., the
Hilbert-Schmidt norm) \cite{gqd1-sp} and the Schatten 1-norm
\cite{gqd2-sp}, which are given, respectively, by
%%%%%%%%%%%%%%%%%%%%%%%%%%%
\begin{equation}\label{eq-m3}
 D_2(\rho)=2\min_{\Pi^A}\parallel\rho-\Pi^A(\rho)\parallel_2^2,
\end{equation}
%%%%%%%%%%%%%%%%%%%%%%%%%%%
and
%%%%%%%%%%%%%%%%%%%%%%%%%%%
\begin{equation}\label{eq-m4}
 D_1(\rho)=\min_{\Pi^A}\parallel\rho-\Pi^A(\rho)\parallel_1,
\end{equation}
%%%%%%%%%%%%%%%%%%%%%%%%%%%
as well as the measurement-induced nonlocality based on the Schatten
2-norm \cite{min1-sp} and the Schatten 1-norm \cite{min2-sp}, which
are given, respectively, by
%%%%%%%%%%%%%%%%%%%%%%%%%%%
\begin{equation}\label{eq-m5}
 N_2(\rho)=2\max_{\Pi^A}\parallel\rho-\Pi^A(\rho)\parallel_2^2,
\end{equation}
%%%%%%%%%
and
%%%%%%%%%%%%%%%%%%%%%%%%%%%
\begin{equation}\label{eq-m6}
 N_1(\rho)=\max_{\Pi^A}\parallel\rho-\Pi^A(\rho)\parallel_1,
\end{equation}
%%%%%%%%%%%%%%%%%%%%%%%%%%%
where $||\varrho||_p=[{\rm Tr}(\varrho^\dag\varrho)^{p/2}]^{1/p}$ is
the Schatten $p$-norm, and $\Pi^A=\{\Pi_k^A\}$ is the set of
projective measurements on subsystem $A$. As $\sum_k (\Pi_k^A\otimes
\mathbb{I}_B) \mathbb{I}_{AB} (\Pi_k^A\otimes \mathbb{I}_B)= \sum_k
(\Pi_k^A\otimes \mathbb{I}_B)^2=\mathbb{I}_{AB}$, we have
%%%%%%%%%%%%%%%%%%%%%%%%%%%
\begin{equation}\label{eq-m7}
 \begin{split}
 ||\rho-\Pi^A(\rho)||_p& = \frac{1}{2}\parallel \chi \hat{n} \cdot\vec{X}-\Pi^A( \chi \hat{n} \cdot\vec{X})\parallel_p,\\
                       & = \frac{1}{2}\chi\parallel\hat{n} \cdot\vec{X}-\Pi^A(\hat{n}\cdot\vec{X})\parallel_p,
 \end{split}
\end{equation}
%%%%%%%%%%%%%%%%%%%%%%%%%%%
and therefore by comparing with Lemma 1, we have $f(\chi)=\chi/2$,
and $g(\hat{n}^s)=g(\hat{n})={\rm opt}_{\Pi^A} \parallel\hat{n}
\cdot\vec{X}-\Pi^A (\hat{n} \cdot\vec{X})\parallel_p$, with ${\rm
opt}\in\{\max,\min\}$.

Similarly, the Hellinger distance discord defined based on the
square root of the density operator $\rho$ \cite{square}, i.e.,
%%%%%%%%%%%%%%%%%%%%%%%%%%%
\begin{equation}\label{eq-m8}
 D_H(\rho)=\min_{\Pi^A}\parallel \sqrt{\rho}-\Pi^A(\sqrt{\rho})\parallel_2^2,
\end{equation}
%%%%%%%%%%%%%%%%%%%%%%%%%%%
also fulfill the requirement of Lemma 1 as $\sqrt{\rho}$ can be
decomposed as $\sqrt{\rho}=\sum_{ij}\gamma_{ij}X_i^A \otimes X_j^B$,
with $\{X_i^A\}_{i=0,1,\ldots,d_A}$ and $\{X_j^B\}_{j=0,1,\ldots,
d_B}$ being the sets of Hermitian operators which constitute an
orthonormal operator bases for the Hilbert space $\mathcal {H}_A$
and $\mathcal {H}_B$ ($d_{A,B}=\dim \mathcal {H}_{A<B}$), and
$\sum_k (\Pi_k^A\otimes \mathbb{I}_B) (X_0^A \otimes X_0^B)
(\Pi_k^A\otimes \mathbb{I}_B) =X_0^A \otimes X_0^B$.

Moreover, when considering the two-qubit states, the maximum
Bell-inequality violation $B_{\rm max}(\rho)$ \cite{Bell}, the
remote state preparation fidelity $F_{\rm rsp}(\rho)$ \cite{rsp},
and $N_{\rm qt}(\rho)$ which is a monotonic function of the average
teleportation fidelity $F_{\rm qt}(\rho) = 1/2+ N_{\rm qt}(\rho)/6$
\cite{qt}, given, respectively, by
%%%%%%%%%%%%%%%%%%%%%%%%%%%
\begin{equation}\label{eq-m9}
  \begin{split}
  & B_{\rm max}(\rho)= 2\sqrt{E_1 + E_2},\\
  & F_{\rm rsp}(\rho)= \frac{1}{2}(E_2 + E_3),\\
  & N_{\rm qt}(\rho) =\sqrt{E_1+E_2+E_3},
  \end{split}
\end{equation}
%%%%%%%%%%%%%%%%%%%%%%%%%%%
also satisfy the requirement of Lemma 1, thus the factorization
relation (8) holds. Here, $E_1\geqslant E_2 \geqslant E_3$ are the
eigenvalues of $T^\dag T$, with $T$ being a $3\times 3$ matrix, and
$T_{ij}={\rm Tr}(\rho\sigma_i\otimes \sigma_j)$, $\sigma_{1,2,3}$
are the usual Pauli operators.

% Non-BibTeX users please use
% Format for Journal Reference

\end{appendix}

\end{document}